\documentclass[lettersize,journal]{IEEEtran}
\usepackage{amsmath,amsfonts, amssymb}
\usepackage{algorithm}
\usepackage[noend]{algpseudocode}
\usepackage{xpatch}
\usepackage{array}
\usepackage{textcomp}
\usepackage{stfloats}
\usepackage{url}
\usepackage{verbatim}
\usepackage{graphicx}
\usepackage{cite}
\usepackage{graphicx}
\usepackage{subcaption} 
\usepackage{caption,subcaption, ragged2e}
\usepackage{xcolor}
\usepackage{balance}

\usepackage{marginnote}
\reversemarginpar

\makeatletter
\xpatchcmd{\algorithmic}{\itemsep\z@}{\itemsep=2ex plus2pt}{}{}
\makeatother

\usepackage[normalem]{ulem}

\begin{document}

\title{EDRP: \underline{E}nhanced \underline{D}ynamic \underline{R}elay \underline{P}oint Protocol for Data Dissemination in Multi-hop Wireless IoT Networks}


\author{Jothi Prasanna Shanmuga Sundaram, Magzhan Gabidolla, Luis Fujarte, Shawn Duong, Jianlin Guo, Toshiaki Koike-Akino, Pu (Perry) Wang, Kieran Parsons, Philip V. Orlik, Takenori Sumi, Yukimasa Nagai, Miguel \'A.\ Carreira-Perpi\~n\'an, Alberto E. Cerpa

\thanks{Jothi Prasanna Shanmuga Sundaram, Magzhan Gabidolla, Luis Fujarte, Shawn Duong, Miguel \'A.\ Carreira-Perpi\~n\'an and Alberto E. Cerpa are with the University of California, Merced (e-mail: \{jshanmugasundaram,mgabidolla, lfujarte , sduong8, mcarreira-perpinan, acerpa\}@ucmerced.edu.}
\thanks{Jianlin Guo, Toshiaki Koike-Akino, Pu (Perry) Wang, Kieran Parsons, Philip V. Orlik are with Mitsubishi Electric Research Laboratories, Cambridge, MA, USA. (e-mail: \{guo, koike, pwang, parsons, porlik\}@merl.com)}
\thanks{Takenori Sumi, Yukimasa Nagai are with 
Information Technology R\&D Center, Mitsubishi Electric Corporation, Kamakura, JAPAN.
(e-mail:\{Sumi.Takenori@dc., Nagai.Yukimasa@ds.\}MitsubishiElectric.co.jp)}

\thanks{Magzhan Gabidolla and Miguel {\'A}.\ Carreira-Perpi{\~n}{\'a}n were partially supported by NSF award IIS--2007147.}

\thanks{Copyright (c) 2026 IEEE. Personal use of this material is permitted. However, permission to use this material for any other purposes must be obtained from the IEEE by sending a request to pubs-permissions@ieee.org.}
}



\maketitle

\begin{abstract}
Emerging IoT applications are transitioning from battery-powered to grid-powered nodes. DRP, a contention-based data dissemination protocol, was developed for these applications. Traditional contention-based protocols resolve collisions through control packet exchanges, significantly reducing goodput. DRP mitigates this issue by employing a distributed delay timer mechanism that assigns transmission-start delays based on the average link quality between a sender and its children, prioritizing highly connected nodes for early transmission. However, our in-field experiments reveal that DRP is unable to accommodate real-world link quality fluctuations, leading to overlapping transmissions from multiple senders. This overlap triggers CSMA’s random back-off delays, ultimately degrading the goodput performance. 

To address these shortcomings, 
we first conduct a theoretical analysis that characterizes the design requirements induced by real-world link quality fluctuations and DRP’s passive acknowledgments.
Guided by this analysis, we design EDRP, which integrates two novel components: (i) Link-Quality Aware CSMA (LQ-CSMA) and (ii) a Machine Learning-based Block Size Selection (ML-BSS) algorithm for rateless codes. LQ-CSMA dynamically restricts the back-off delay range based on real-time link quality estimates, ensuring that nodes with stronger connectivity experience shorter delays. ML-BSS algorithm predicts future link quality conditions and optimally adjusts the block size for rateless coding, reducing overhead and enhancing goodput. In-field evaluations of EDRP demonstrate an average goodput improvement of 39.43\% than the competing protocols.

\end{abstract}


\section{Introduction}
\label{sec:intro}

\IEEEPARstart{E}{merging} IoT applications, such as peer-to-peer (P2P) energy trading in smart homes, smart meters, building management, and smart grids, are increasingly adopting grid-powered nodes \cite{yang2024rateless, nagai2024improve, sundaram2024mars}. This represents a departure from traditional IoT deployments, which typically rely on battery-powered nodes, as seen in applications like  irrigation monitoring \cite{winkler2018irrigation}. In these emerging scenarios, the IoT radio constitutes only a minor component of the overall device, which is predominantly powered by the grid or supported by a larger battery reserve \cite{nagai2024improve, sundaram2024mars, yang2024rateless} compared to the limited battery capacities in conventional IoT devices. This paradigm shift, along with the advent of these novel IoT applications, highlights the need for advanced protocols for nextG IoT apps.

Data dissemination protocols \cite{arakadakis2021firmware, sundaram2025machine, du2016pando} for multi-hop IoT networks have been heavily studied 
because of their important applications like firmware updates. Data dissemination protocols for multi-hop IoT networks can be categorized into (i) contention-based and (ii) contention-free protocols. Contention-based 
protocols \cite{arakadakis2021firmware, sundaram2025machine} employ CSMA MAC to break the tie between multiple senders contending for transmission. These protocols cannot use spatial reuse and heavily suffer from back-off delays. Contention-free protocols \cite{du2016pando} employ TDMA and Constructive Interference (CI) to improve transmission reliability. The latter requires a tight time synchronization of 0.5 $\mu$s between multiple sender nodes to generate concurrent transmissions \cite{zimmerling2017adaptive}, 
while recent time synchronization protocols \cite{sarvghadi2016message}, like DESAA \cite{sarvghadi2022distributed}, can achieve an average time-sync error of $\pm$30.01$\mu$s - $\pm$67.81$\mu$s.
Although contention-free protocols perform better than contention-based protocols, they are seldom used in emerging IoT applications \cite{jokar2016intrusion, guo2023ieee} as it is tough to maintain tight time-sync 
required for CI-based contention-free data dissemination protocols.

The latest time-sync protocol \cite{Huan2022Microsecond} presents a microsecond level time synchronization method for LoRa networks using low layer timestamping and an asymmetric time translation framework, achieving high accuracy through two way communication between nodes and the gateway. Another recent work \cite{Huan2023OneWay} builds on this approach by introducing a one way reverse asymmetric framework, which eliminates the need for uplink responses from nodes. This improvement significantly reduces energy consumption, lowers protocol complexity, and enhances scalability. Together, these works advance practical, energy efficient synchronization for long lifetime LoRa IoT deployments without sacrificing precision. However, these methods, developed for single hop LoRa networks, are not directly applicable for multi-hop Zigbee networks.

The emerging IoT applications are mostly deployed in an urban/semi-urban environment in the mesoscale range \cite{sundaram2024mars} of 0.1-1Km. The networking performance of IoT deployments in these environments is affected by (i) environmental obstacles and (ii) vagarious IoT radios. These environments consist of buildings made of different materials such as wood, glass, firewalls, steel, etc. These environments also consist of human influx and vehicle movements. These environmental obstacles attenuate wireless signals erratically. Moreover, IoT radios are designed to be cheaper than traditional radios, such as LTE/5G/Wi-Fi.
So, the IoT radios are vagarious. For example, an IoT radio may deviate slightly from the configured central frequency during transmission~\cite{sundaram2019survey}. These characteristics make it hard to achieve any time-sync efforts \cite{zimmerling2020synchronous} for contention-free data dissemination protocols.   

The contention-free protocols do not use any control packets as they use TDMA-based MAC protocol. In TDMA MAC, each node is allotted a time slot to transmit packets using MAC-level synchronization packets, avoiding the exchange of control packets by the data dissemination protocols. On top of this improvement, multiple, tightly time-synchronized senders constructively interfere\cite{zimmerling2017adaptive} to improve transmission reliability. However, the above-discussed deployment environments and characteristics of the emerging IoT applications are a hindrance to contention-free protocols. A major drawback of the contention-based protocols \cite{arakadakis2021firmware, sundaram2025machine} is the control packet exchanged by the data dissemination protocol and their back-off delays. In the case of multiple potential senders, the nodes in contention-based protocols exchange control packets to choose one potential sender. During the exchange of these control packets, CSMA is used, leading to back-off delays. 

First, we conducted an in-field evaluation of DRP \cite{guo2023rateless}. DRP is the recently developed contention-based data dissemination protocol that avoids the exchange of control packets through distributed delay timers. Since they use rateless codes to disseminate data, they avoid the exchange of ACK for rateless codes using a passive acknowledgment scheme (more details in 
Section
\ref{sec:related}). However, our in-field performance analysis of DRP did not achieve the expected results (more details in 
Section
\ref{sec:DRP_perf}). This performance degradation is because the link quality models of the simulators do not replicate the link quality fluctuations happening in real-world scenarios~\cite{boano2018iot}. DRP's dynamic delay timer mechanism cannot handle real-world link quality fluctuations. So, the DRP's delay timer is not able to isolate the transmissions of multiple senders to avoid collisions and backoff delays, leading to a degradation in DRP's goodput performance. To solve this problem, we took a straw-man approach and did a grid search to find the right values for the delay timer variables of DRP. This intuitive optimization did not ameliorate the performance of DRP.

Second, we developed Enhanced-DRP (EDRP) to solve the aforementioned 
problems. EDRP comprises two components: (i) Link Quality-aware CSMA (LQ-CSMA), and (ii) ML-based Block Size Selection (ML-BSS) for rateless codes. LQ-CSMA is aware of the real-world link quality fluctuations while assigning transmission-start delays for potential senders. It prioritizes highly connected nodes for early transmission. Our in-field experiments showed that link quality erratically changes over time. Setting smaller and bigger block sizes for rateless codes during good and bad link-quality conditions respectively, decreases goodput. So, we predict the future link quality conditions using TAO\cite{carreira2018alternating}-optimized ordinal regression trees  to set appropriate block sizes. In-field experiments show that EDRP improves goodput by 39.43\% than its competitors.

In summary, our contributions are:
\begin{enumerate}
    \item Through in-field experiments of DRP, we identify the key factors degrading its performance and show that the DRP's distributed delay timer mechanism is unable to robustly handle erratic link-quality fluctuations in real-world deployments. We further provide a theoretical characterization of the resulting design constraints and trade-offs governing collision behavior and dissemination efficiency. 
    \item We designed and developed LQ-CSMA to overcome the aforementioned factor affecting the performance of DRP.
    \item We developed ML-BSS powered by TAO\cite{carreira2018alternating}-optimized decision tree that predicts future link quality conditions to optimize block sizes for rateless codes.
\end{enumerate}

\vspace{-0.15in}
\section{Related Work}
\label{sec:related}

Data dissemination protocols for multi-hop IoT networks can be categorized into (i) contention-based \cite{arakadakis2021firmware, sundaram2025machine} and (ii) contention-free \cite{ du2016pando} protocols. In this article, we focus on contention-based protocols that use CSMA to break ties between multiple senders 
as emerging IoT devices are resource constrained and cannot satisfy the time-sync requirements of contention-free data dissemination protocols (e.g., Glossy \cite{zimmerling2017adaptive}, Pando \cite{du2016pando}), which requires microsecond-level synchronization between transmitters to enable concurrent transmissions, such synchronization can be difficult to maintain in emerging IoT environments characterized by building-induced signal attenuation, dynamic interference. Although there are several protocols \cite{sarvghadi2016message} like DESAA \cite{sarvghadi2022distributed} that achieves efficient scheduling with relaxed synchronization requirements.

Rateless Deluge \cite{arakadakis2021firmware} is a controlled flooding protocol. A sender possessing the data to be transmitted will advertise the availability. Receiver(s) that need \textit{x} number of blocks to decode the data waits for a random back-off period to see if there are any other receiver(s) that send a request for \textit{y} number of blocks, where $y \geq x$, to avoid redundant requests. A sender will wait for a certain period and broadcast the highest requested number of blocks.
 
MNP \cite{arakadakis2021firmware} reduces collisions by electing a sender per neighborhood. A sender is elected based on their ability to successfully reach a higher number of nodes. When a transmitting node has all the data to be transmitted, it sends an advertisement packet. The nodes that need this data will send back a reply. These request and reply packets are utilized to calculate how many successful neighbors a sender can reach. If there are multiple potential senders in the neighborhood, they overhear these advertisement-reply packets and identify which sender can reach a higher number of neighbors.
If a potential sender finds another node with a higher number of neighbors, 
it goes to sleep, letting the other node transmit.

AdapCode \cite{sundaram2025machine} employs rateless coding for data dissemination. They identify through experiments that for high-density neighbor scenarios, a higher degree ($d_n$) leads to higher reliability. So, they develop a hard-coded degree distribution for different neighbor density scenarios and utilize this degree ($d_n$) to encode packets during data dissemination.

DRP \cite{guo2023rateless} avoids the exchange of control messages through a distributed, dynamic timer based on the Link Quality (LQ) of the sender with its children. An important requirement of data-dissemination protocols is to disseminate bulk data to all the nodes in the network as quickly as possible. When there are multiple potential senders, the distributed dynamic timer of DRP \cite{guo2023rateless} breaks the tie without any exchange of control messages as follows:  A node that is connected to a higher number of children with high-quality links is assigned the lowest delay time to start transmissions. This can be calculated based on the local LQ between a sender and its children, without exchanging any control messages. As the highly connected nodes with high-quality links to their children are given a lower delay, this helps fast dissemination of data to the entire network.  All the contention-based data-dissemination approaches use rateless coding to disseminate data. In rateless coding, the sender encodes the data and sends it to the receiver(s) continuously until the receiver(s) sends an ACK back to the sender. DRP avoids this by using a passive acknowledgment mechanism: Once the receiver decodes the data, it starts transmitting it to its children. This transmission heard by the sender is considered an ACK. When the sender has multiple children nodes, it stops transmitting once it receives an ACK from a certain percentage of its children. They also state that ARQ is more efficient during good link quality scenarios than rateless coding \cite{guo2023rateless}. So, they switch between ARQ and rateless based on the current link quality estimation. NS3 simulations indicate that DRP enhances effective throughput by an average of 604\%.
\begin{figure}[t]
    \centering
     \includegraphics[width=0.9\linewidth, height=7cm]{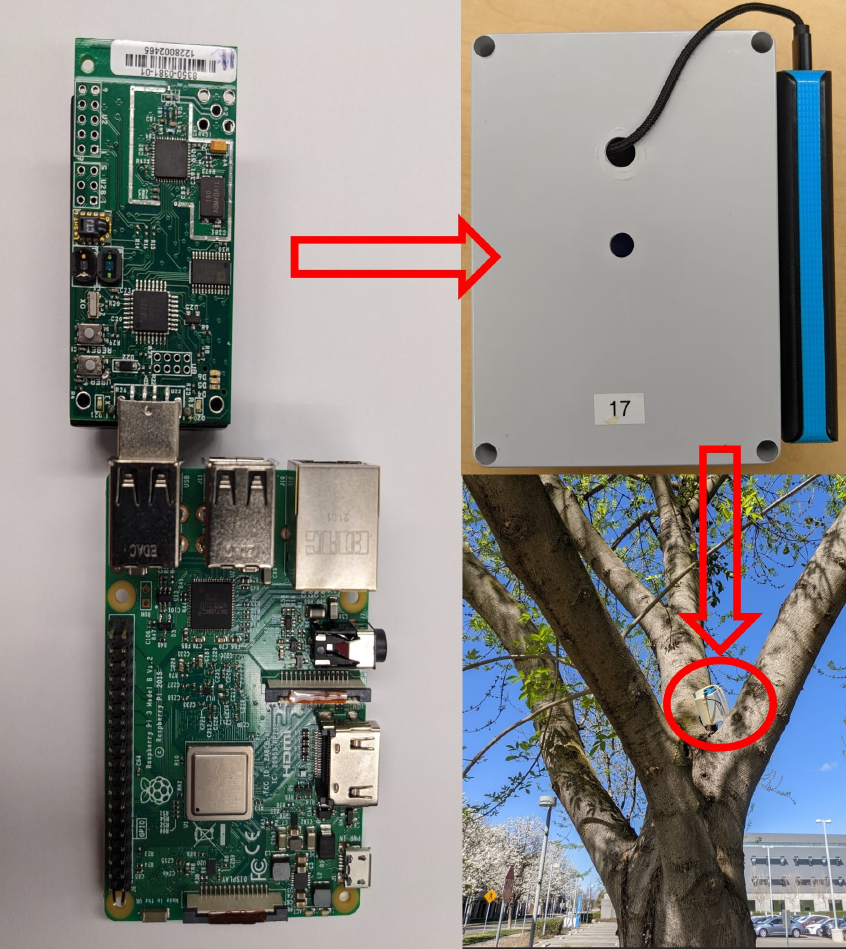} 
     \caption{Hardware setup}
    \label{fig:hardware}   
\end{figure}

\vspace{-0.15in}
\section{Real-world performance analysis of DRP}
\label{sec:DRP_perf}

The above-stated results of DRP \cite{guo2023rateless} were obtained only through NS3 simulations. In contrast, we aim to evaluate its performance in real-world deployment scenarios. 

\begin{figure}[t!]
\centering
    \includegraphics[width=0.9\linewidth, height=7cm]{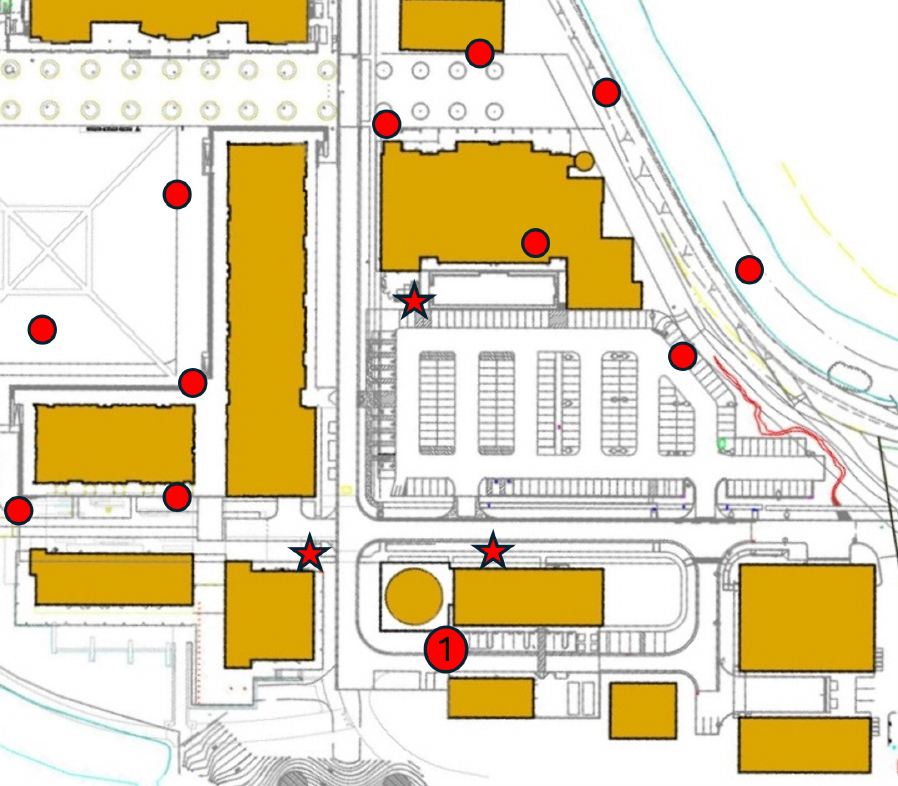} 
    \caption{Topology setup}
    \label{fig:topo}
\end{figure}

\textbf{\textit{Experimental settings.}} We conducted experiments using the hardware setup shown in Figure \ref{fig:hardware}. The TelosB \cite{orbitlabHardwarefDeviceskTinyOSx2013} Zigbee mote is plugged into a Raspberry Pi 3B powered by an external power bank. This hardware setup is enclosed in a PVC casing during deployment. The log files are stored in the Raspberry Pi 3B. We set up 15 
static, non-mobile
nodes as shown in Figure \ref{fig:topo}. The TelosB mote is programmed using TinyOS \cite{sundaram2025comnets}. The nodes form a mesh network using a simple distance-vector protocol that mimics RPL \cite{lamaazi2017rpl} DRP runs for 100 iterations. Node-1 disseminates 1000 bytes of data into the network during each iteration. The "star" nodes are the children of Node-1. The rest of the "circle" nodes communicate with node-1 through the "star" nodes.

\textbf{\textit{Benchmarks.}} DRP\cite{guo2023rateless} is compared with three different contention-based data dissemination protocols MNP~\cite{arakadakis2021firmware}, Rateless deluge~\cite{arakadakis2021firmware}, and AdapCode~\cite{sundaram2025machine}, discussed in 
Section
\ref{sec:related}.

\textit{\textbf{Metric.}} Goodput is the metric employed to compare these different protocols. Goodput is the number of useful bits successfully transmitted per unit of time. We chose goodput instead of throughput because goodput does not account for overhead data such as packet headers, trailers, and CRC bytes. 

\begin{figure}[]
    \centering
    \includegraphics[width=0.9\linewidth, height=7cm]{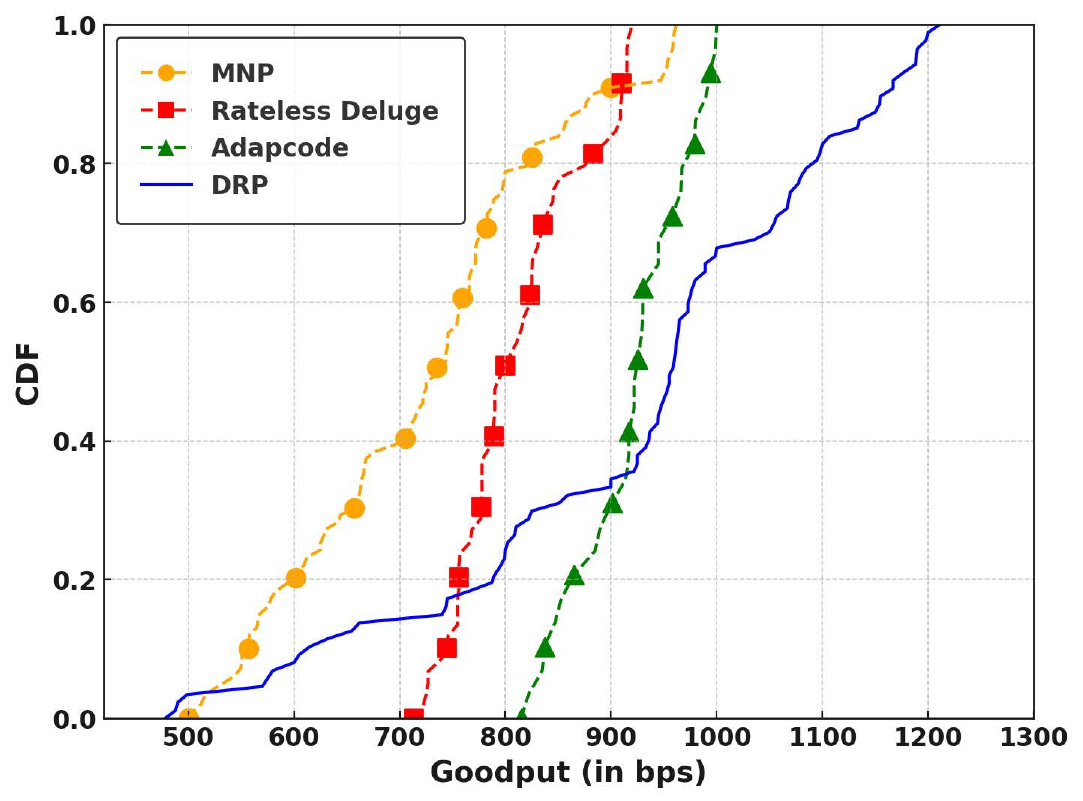}
    \caption{Goodput of DRP}
    \vspace{-0.6cm}
    \label{fig:mot_cdf}
\end{figure}

\textbf{\textit{Result Analysis.}} The goodput of these protocols is plotted as 
Cumulative Distribution Function (CDF)
in Figure \ref{fig:mot_cdf}. 
This graph shows that MNP achieves the lowest goodput and DRP achieves the highest goodput. DRP achieves an average goodput gain of 27.87\%, 13.80\%, and 0.46\% over MNP, Rateless Deluge, and AdapCode, respectively. Overall, DRP achieves a marginal average goodput gain of  12.94\% than the other competing protocols. Until 38\% of the transmissions, DRP does not achieve any goodput gain than the competing protocol, AdapCode \cite{sundaram2025machine}. 

It should be noted that DRP has a long-tail start, indicating poor performance of DRP for these transmissions. The performance of DRP plummets due to the inefficiency of DRP delay timers to isolate the transmission of multiple senders in the same neighborhood.  From 38\%-64\% of the transmissions, DRP shows a limited goodput gain.
From 64\%-100\% of the transmissions, DRP shows marginal goodput gains over the other protocols. 
The above-observed goodput gain of DRP is relatively modest when compared with other protocols in the literature. This raises concerns regarding its practical applicability in real-world deployments where substantial performance gains are required for adoption. This observation motivates a deeper investigation into the underlying factors contributing to the performance limitations of DRP, intending to identify and address potential areas for performance improvement.

\section{Insights into the performance of DRP}
\label{sec:insights}
\begin{figure*}[t]
    \centering
    \includegraphics[width=\textwidth, keepaspectratio]{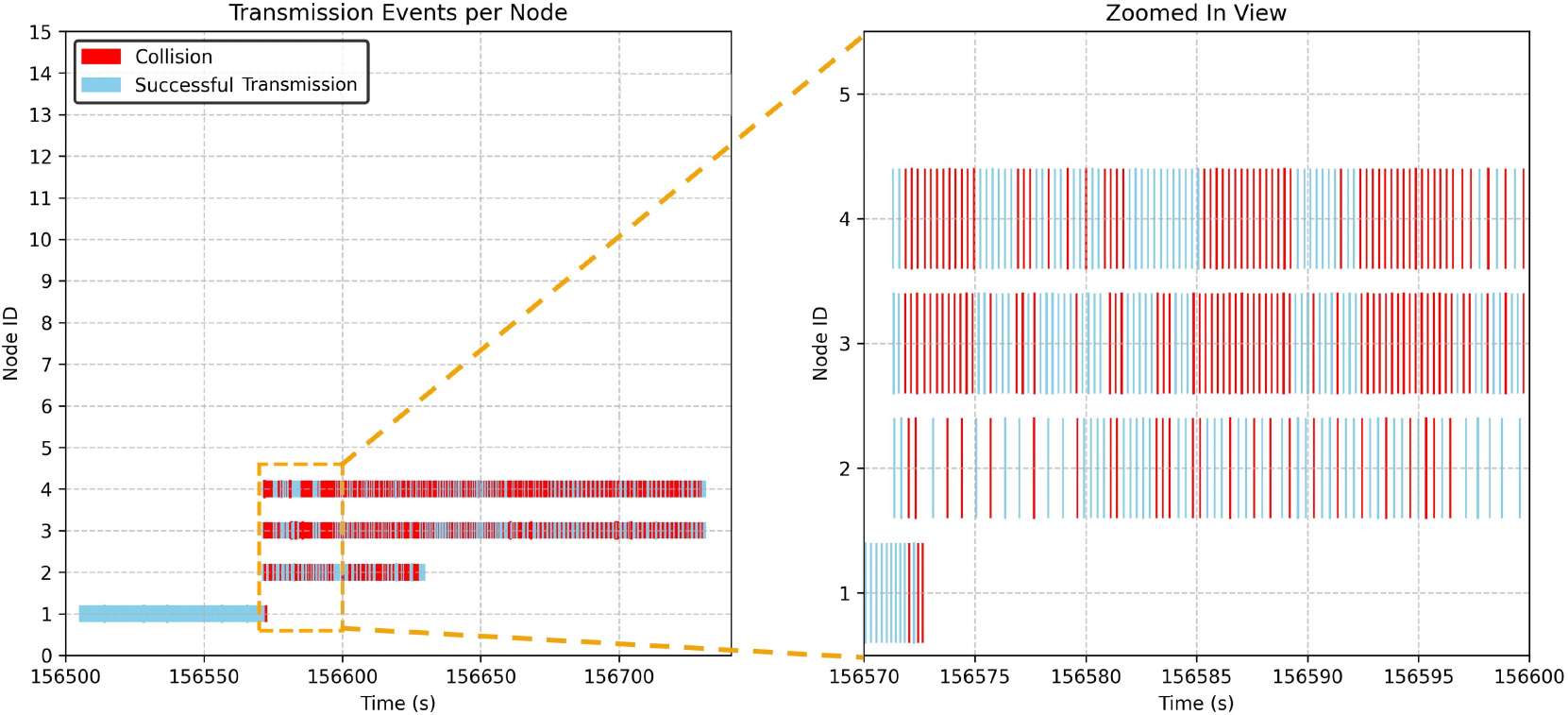}
    \caption{Transmission events showing successful and collided transmissions of multiple senders in the same neighborhood}
    \label{fig:events}
\end{figure*}

DRP \cite{guo2023rateless} eliminates the need for control packet exchanges to start transmissions by assigning dynamic delay times to potential senders. This delay time is calculated based on the link quality between the sender and its children. This approach is theoretically expected to deliver significant goodput gains, as it does not rely on control packets for sender selection. However, real-world experiments, as discussed in the previous section, do not fully reflect these anticipated benefits. To investigate this discrepancy, we performed an in-depth analysis on the transmission events of DRP. In this section, we present key findings from this in-depth analysis. 

To facilitate comprehension, a representative excerpt of the transmission events for various transmitting nodes is visualized in Figure~\ref{fig:events}. In this excerpt, we discuss only the transmissions of Node-1 and its direct children, "star" nodes (see Fig. \ref{fig:topo}).  On the left of Figure \ref{fig:events}, we depicted the transmission events of the root node (Node-1) and their children node. For further details, a short time interval denoted by the golden-box is shown in a magnified view on the right side of Figure \ref{fig:events}.

We time-stamped the transmission and reception of the packets at a micro-second level of granularity and compared them offline to plot this graph. In Figure~\ref{fig:events}, the blue lines denote successful transmission, without any collision and the red lines denote collision. On the left-hand side, we can see that the graph starts with a long sequence of blue lines for Node 1. At this time, Node-1 (Base station) is the only transmitting node. At the beginning of the "yellow box" on the left-hand side figure, Node 1 stops transmitting slightly after nodes 2,3,4 starts transmitting data to their children, denoting the working of the passive ACK scheme of DRP\cite{guo2023rateless}. It is also seen that the transmission of nodes 2,3,4 is close in the figure, but the logs say that they are separated only by a few milliseconds, while it is apparent that more time is needed between each node to successfully transmit 1000 bytes of data.  


Upon conducting an in-depth analysis of the logs, we found that the link quality of each of these star nodes with their respective children nodes is similar. This link quality is used as the input to the delay-timer of DRP at each node, leading to a similar delay time for all these nodes. This phenomenon is seen to be repeated in the log files, for $\approx$80\% of the disseminations. This is the reason for the poor performance improvement of DRP\cite{guo2023rateless}, regardless of the theoretically anticipated performance improvement that should stem from the eradication of control packets. The transmission of nodes 2,3,4 should be completed within a short period if they are isolated far enough by the DRP delay timer. However, the contrary is observed because of the collisions. 

\begin{figure}[t]
    \centering
     \includegraphics[width=0.9\linewidth, height=6.5cm]{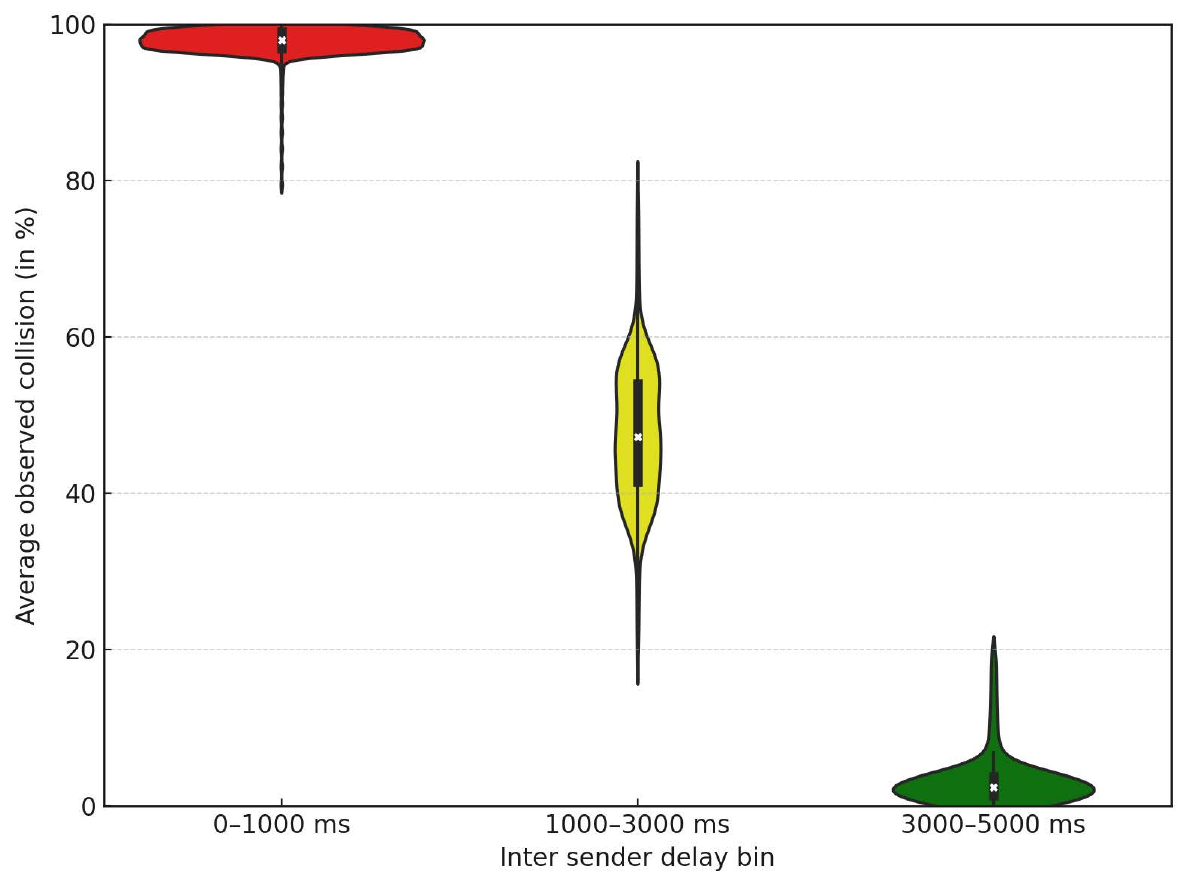} 
     \caption{Technical analysis on the effect of inter sender delay on the collisions during disseminations. }
    \label{fig:tech_isd}   
\end{figure}

We conducted a technical analysis to understand the effect of inter sender delay on collisions. Figure \ref{fig:tech_isd} plots the collision rate as the function of inter sender Delay between potential senders in the same collision domain. We performed a detailed empirical analysis of dissemination traces from the in-field deployment to study how the inter-sender delay gap impacts the observed collision rate. For each dissemination round, we measured the delay gap between senders in the same collision domain and grouped these into 3 delay regions: (i)0–1000 ms,(ii)1000–3000 ms, (iii)3000–5000 ms.

We then computed the observed collision rate in each region as the fraction of packets lost due to collision, derived directly from transmission logs. These transmission logs reveal that DRP frequently schedules senders within the 0–1000ms region, where the collision rate is consistently near 100\% with almost no variability, highlighting DRP's failure to ensure adequate delay separation when multiple nodes have similar link quality. The middle region (1000–3000ms) shows partial improvement but still exhibits considerable variation in collision outcomes. Only in the 3000–5000ms region we see a consistent drop in collision rates to below 5\%, indicating that delay gaps in this range are safe for sender isolation. This technical analysis demonstrates that DRP's delay timers, which depend on Link Quality (LQ), often assign similar delay values to multiple nodes, particularly when many nodes have similar LQ estimates. As a result, DRP fails to sufficiently separate concurrent transmissions, leading to collisions.

The right-hand side of Figure \ref{fig:events} shows a magnified view of the events for a smaller period denoted by the golden box. Here, we can see that the blue and red lines denoting transmissions have varying intervals between them. The transmission events of Node 2 are sparse while the transmission events of Node 3 and 4 are denser than the events of Node 2. We can also see that the events of Node 2 are sometimes dense and sometimes sparse. This is due to the CSMA's random back-off delay for each transmission.  During the dissemination of 1000~bytes of bulk data, this data is broken into multiple packets before transmission, and these broken packets are reordered at the receiver to access the bulk data. DRP \cite{guo2023rateless} schedules only the beginning of the transmission of bulk data and it does not affect the transmission of each packet during this bulk data dissemination. These packets experience random CSMA back-off delays, leading to DRP's poor performance. 

\begin{figure}[t]
    \centering
     \includegraphics[width=0.9\linewidth, height=6.5cm]{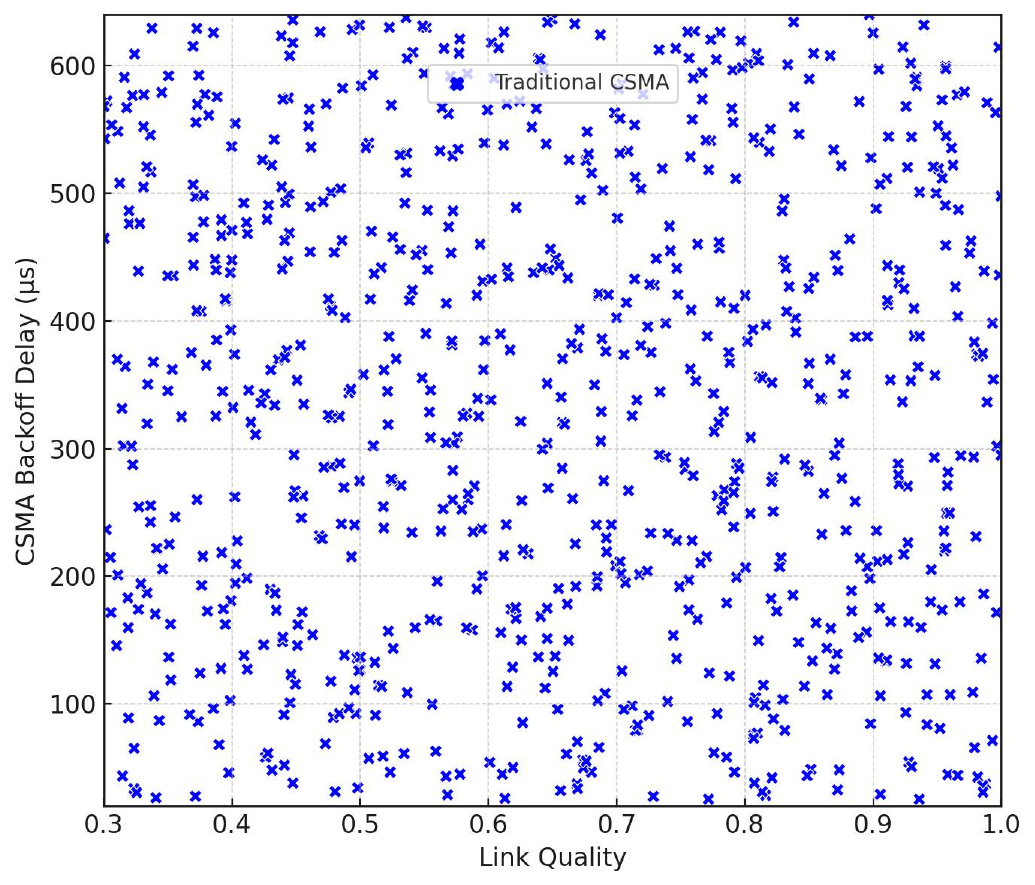} 
     \caption{Technical analysis on the back-off delay experienced by packets under traditional CSMA.}
    \label{fig:tech_csma}   
\end{figure}

We conducted a technical analysis to validate this statement. For this analysis, we plotted Figure \ref{fig:tech_csma}, plotting CSMA backoff delay (Y-axis, in ms) experienced be packet as the function of the corresponding LQ (X-axis) estimated. This figure shows that the resulting distribution is uniform and uncorrelated, with no trend suggesting that CSMA follows the principle of DRP to assign lower delays to high quality links. In fact, the Pearson correlation coefficient \cite{benesty2018pearson} was -0.027, very close to zero, confirming that CSMA assigns delays randomly as expected, irrespective of the link quality. This reinforces our claim that even if DRP's delay timers are imperfect but deterministic, the randomized nature of CSMA further negates any prioritization intent, leading to increased collisions when delays overlap.

The \textbf{key insights} from this performance analysis are:
\begin{enumerate}
    \item The delay timer of DRP\cite{guo2023rateless} does not provide the theoretically anticipated performance improvement in real-world deployments because of the similar link quality experienced by senders in same neighborhood. 
    \item The above-discussed link quality phenomenon heavily degrades the performance of DRP and is repeatedly found in the logs, $\approx$80\% of the disseminations. 
    \item DRP schedules the beginning of bulk data dissemination but it does not affect the transmission of multiple different packets during data dissemination. 
    \item These packets experience CSMA's random back-off delays, degrading the performance of DRP.
\end{enumerate}

While the above analysis focuses on the limitations of DRP’s delay-based sender selection and its impact on collision behavior, an additional challenge arises from the use of rateless coding under a passive acknowledgment scheme of DRP. In DRP, senders do not receive explicit feedback regarding decoding progress or packet corruption at receivers, making online adaptation of coding parameters infeasible. Prior work \cite{du2016rateless} has shown that rateless block size selection significantly affects the trade-off between protocol overhead and loss sensitivity; however, existing adaptation mechanisms \cite{yang2024rateless} rely on receiver feedback and are therefore incompatible with DRP. As a result, block size selection in DRP remains static and potentially suboptimal under fluctuating link conditions, motivating a predictive block size adaptation approach that operates without receiver feedback. These unresolved constraints motivate a formal characterization and analysis of the underlying design requirements and trade-offs, which we present in Section \ref{sec:theory}.

\vspace{-0.10in}
\section{Theoretical Analysis of Design Requirements}
\label{sec:theory}
In this section, we formally characterize the requirements and trade-offs that any protocol operating under the passive acknowledgments and fluctuating link qualities must satisfy, thereby guiding the design of EDRP presented in Section \ref{sec:edrp-design}.

\vspace{-0.10in}
\subsection{Analysis of Inter-Sender Delay and Collision Probability}
In this subsection, we formally establish the monotonic dependence of collision probability on inter-sender delay separation. We first characterize how overlap among bulk data dissemination rounds induces packet-level contention, and then show that increasing the effective delay separation among competing packet transmissions strictly reduces packet-level collision probability.

Bulk data dissemination in DRP occurs in rounds rather than as single atomic transmissions. Let $t_i^{(r)}$ denote the start time of dissemination round $r$ at sender $i$, $t_j^{(r')}$ denote the start time of dissemination round $r'$ at sender $j$, sender $i$ and sender $j$ are two competing senders in the same collision domain, i.e., they can interfere with each other. Let $T_r$ denote the duration of a dissemination round. These two rounds overlap if
\begin{equation}
|t_i^{(r)} - t_j^{(r')}| < T_r .
\end{equation}
Here, $r$ and $r'$ index dissemination rounds initiated independently by different senders and need not be synchronized. 

Dissemination rounds may overlap when multiple potential senders exhibit similar link-quality characteristics. In such cases, link-quality-based sender differentiation is insufficient to induce strict temporal isolation among competing transmitters. As a result, multiple senders may initiate bulk dissemination rounds within comparable time intervals, leading to partial or complete overlap of their transmission durations.

Each dissemination round consists of a sequence of packet transmissions rather than a single atomic transmission. Let
\begin{equation}
T_r = \sum_{k=1}^{N_r} T_p^{(k)},
\end{equation}
where $N_r$ denotes the number of packets transmitted in the round and $T_p^{(k)}$ denotes the effective service time of the $k$-th packet transmission, which includes both the physical packet transmission duration and the CSMA-induced backoff delay incurred prior to channel access. Since CSMA backoff durations are random and protocol-dependent (e.g., IEEE~802.15.4 or IEEE~802.11), $T_p^{(k)}$ is modeled using its expected value in the theoretical analysis. This modeling choice captures average packet-level contention behavior while preserving analytical tractability and generality. When dissemination rounds overlap, packets from different rounds interleave on the channel and contend independently under CSMA. Consequently, dissemination round-level overlap induces packet-level contention, and collision behavior is governed by the CSMA for individual packet transmissions.

Consider two packet transmissions with start times $t_i$ and $t_j$, respectively. The inter-sender delay separation is defined as
\begin{equation}
\Delta \stackrel{\mathrm{def}}{=} |t_i - t_j|.
\end{equation}

Let $T_p$ denote the packet transmission duration. A collision occurs if the first packet transmission has not completed before the second sender starts, i.e., when $\Delta < T_p$. The collision probability conditioned on $\Delta$ can therefore be written as
\begin{equation}
\label{eq:pc_delta}
P_c(\Delta) = \mathbb{P}(T_p > \Delta) = 1 - F_{T_p}(\Delta),
\end{equation}
where $F_{T_p}(\cdot)$ denotes the cumulative distribution function (CDF) of $T_p$. Here, $T_p$ is modeled as a random variable representing packet service time. By definition, $F_{T_p}(\Delta)=\mathbb{P}(T_p \le \Delta)$ is the probability that a packet transmission (including CSMA back-off) completes within time $\Delta$. A collision occurs when the transmission is still ongoing at the time the second sender begins transmission, i.e., when $T_p > \Delta$, yielding $P_c(\Delta)=1-F_{T_p}(\Delta)$. The CDF formulation is introduced solely to enable a formal monotonicity argument with respect to the delay separation $\Delta$, without assuming any specific distribution for $T_p$.

Taking the first-order derivative of equation \eqref{eq:pc_delta} with respect to $\Delta$ yields
\begin{equation}
\frac{dP_c(\Delta)}{d\Delta} = -f_{T_p}(\Delta) < 0,
\end{equation}
where $f_{T_p}(\Delta)$ denotes the probability density function of $T_p$, defined as the derivative of its CDF. Since $f_{T_p}(\cdot)$ is a probability density function, it is non-negative everywhere, i.e., $f_{T_p}(\Delta) \ge 0$ for all $\Delta$, which establishes the strict monotonicity. 

In multi-sender collision domains, collision behavior is dominated by the minimum delay separation among competing transmissions, defined as
\begin{equation}
\Delta_{\min} \stackrel{\mathrm{def}}{=} \min_{i \neq j} |t_i - t_j| .
\end{equation}
Since a collision event occurs when $\exists\, (i,j)$ such that $|t_i - t_j| < T_p$, the probability of collision depends on whether $\Delta_{\min} = \min_{i \neq j} |t_i - t_j|$ falls below the packet duration, making $\Delta_{\min}$ the dominant factor in multi-sender collision behavior.

Reducing collision probability therefore requires increasing $\Delta_{\min}$ in expectation. This can be achieved by enforcing an ordering on channel access delays such that
\begin{equation}
\label{eq:bo_ordering}
\mathbb{E}[T_{bo} \mid LQ_{\text{high}}] < \mathbb{E}[T_{bo} \mid LQ_{\text{low}}],
\end{equation}
where $T_{bo}$ denotes the random backoff duration prior to channel access. The expectation operator $\mathbb{E}[\cdot]$ represents the average backoff duration, while the conditional notation indicates that the expectation is taken given a particular link-quality class. The events $LQ_{\text{high}}$ and $LQ_{\text{low}}$ denote abstract high- and low-link-quality conditions, respectively, without assuming any specific measurement or estimation mechanism.

The inequality in \eqref{eq:bo_ordering} represents a deliberate ordering of channel access delays based on relative link quality, such that senders with higher-quality links experience shorter expected backoff durations, while lower-quality links defer longer on average. This asymmetric ordering increases the expected temporal separation between competing packet transmissions, thereby mitigating packet-level collisions arising from overlapping dissemination rounds.

\vspace{-0.10in}
\subsection{Theoretical Analysis of Rateless Block Size Selection}

We analyze rateless block size selection by explicitly accounting for the structure of bulk data dissemination, the separation between packet-level transmission and block-level decoding, and the use of block-level CRC under the passive acknowledgment scheme of DRP. The purpose of this analysis is to formally characterize the structural trade-offs induced by block size selection and to motivate adaptive mechanisms under stochastic wireless channel conditions.

Bulk data dissemination involves transferring data objects that are significantly larger than the maximum packet payload supported by the underlying PHY/MAC layer. As a result, bulk data must be fragmented and transmitted across multiple packets. The bulk data is first partitioned into encoded rateless blocks of size $B$ bytes, which represent the logical units for decoding. Multiple encoded blocks are packed into the payload of each packet and transmitted over the wireless channel.

Under rateless coding, packet-level CRC checks are disabled and decoding correctness is enforced at the block level using independent CRCs attached to each encoded block. Consequently, a packet that is partially corrupted may still carry usable blocks, while only those blocks whose CRC checks fail are discarded. A block is considered successfully delivered only if its CRC passes, and loss is therefore defined at the block level rather than at the packet level. under the passive acknowledgment scheme of DRP, the sender does not receive explicit feedback regarding packet corruption or block decoding outcomes. The achieved goodput under block size $B$ can be expressed as
\begin{equation}
G(B) = \frac{\mathbb{E}[U(B)]}
{\mathbb{E}[U(B)] + \mathbb{E}[O(B)] + \mathbb{E}[L(B)]},
\end{equation}
where $\mathbb{E}[U(B)]$ denotes the expected amount of useful data successfully decoded at the receiver, $\mathbb{E}[O(B)]$ denotes the expected protocol overhead, and $\mathbb{E}[L(B)]$ represents the expected wasted transmission effort due to blocks that fail CRC checks and are discarded. Expectations represent average behavior over stochastic wireless channel realizations.

Block size induces a fundamental trade-off between overhead efficiency and loss amplification. Each encoded block carries its own CRC. As the block size increases, the number of blocks required to represent a given bulk data object decreases, reducing the total CRC overhead. This implies
\begin{equation}
\mathbb{E}[O(B)] \propto \frac{1}{B},
\end{equation}
indicating that larger block sizes amortize overhead efficiently.

Conversely, increasing the block size aggregates a larger amount of data into a single decoding unit. Under block-level CRC protection, decoding success is binary at the block level: a block is either fully accepted or entirely discarded. While corrupted packets may still contribute usable blocks, failure to decode a block results in all data associated with that block being wasted. For any non-zero probability of block decoding failure under stochastic wireless channel conditions, the expected wasted transmission effort therefore increases with block size. To formally capture this monotonic dependence, we consider the first-order derivative of the expected loss with respect to block size:
\begin{equation}
\frac{d\,\mathbb{E}[L(B)]}{dB} > 0.
\end{equation}
This derivative is strictly positive because, for any non-zero probability of block decoding failure, increasing the block size increases the amount of data discarded upon a failure, thereby increasing the expected wasted transmission effort.

The opposing monotonic behaviors of $\mathbb{E}[O(B)]$ and $\mathbb{E}[L(B)]$ imply that the goodput function $G(B)$ is generally non-monotonic in $B$ and admits an optimal operating point that depends on the underlying link condition. Let $\theta$ denote the (unobserved) channel state during a dissemination round, and define the optimal block size under $\theta$ as
\begin{equation}
B^*(\theta) \stackrel{\mathrm{def}}{=} \arg\max_{B} G(B \mid \theta).
\end{equation}

under the passive acknowledgment, the sender does not observe $\theta$ and cannot directly infer $B^*(\theta)$. Selecting a block size $B \neq B^*(\theta)$ therefore incurs a performance loss, which can be quantified using regret:
\begin{equation}
\mathcal{R}(B,\theta) \stackrel{\mathrm{def}}{=} G(B^*(\theta) \mid \theta) - G(B \mid \theta).
\end{equation}
Here, regret is used strictly as a performance gap metric. 

Since channel conditions vary over time and remain unobserved at the sender, any fixed block size selection strategy incurs non-zero expected regret:
\begin{equation}
\mathbb{E}[\mathcal{R}(B,\theta)] > 0.
\end{equation}
Reducing this expected regret requires adapting block size selection using sender-side observations correlated with packet-level transmission outcomes.

Let $\mathcal{B} = \{B_1, B_2, \dots, B_K\}$ denote the finite set of available rateless block sizes, ordered such that $B_1 < B_2 < \dots < B_K$. The block size selection problem therefore exhibits an inherent ordinal structure: selecting a block size closer to the optimal incurs a smaller regret than selecting one that is farther away. Consequently, the objective is not to predict an exact continuous value nor to classify unrelated categories, but to infer the relative ordering of candidate block sizes under varying channel conditions. This structural property naturally motivates ordinal prediction models for block size selection.

These theoretical observations establish rateless block size as the primary sender-controlled variable under the passive acknowledgment and motivate adaptive, order-aware block size selection mechanisms, described in Section \ref{sec:edrp-design}.

Taken together, the above analysis identifies two fundamental design requirements for efficient bulk data dissemination under the passive acknowledgment scheme and fluctuating link qualities: (i) packet-level collision probability must be reduced by increasing effective inter-sender delay separation among competing transmissions, and (ii) rateless block size must be adaptively selected to balance overhead efficiency against loss amplification under stochastic channel conditions. These requirements arise directly from the structural properties of dissemination rounds, CSMA-based packet contention, and block-level decoding semantics, rather than from any specific protocol realization directly guiding EDRP's design. 

\section{EDRP: \underline{E}nhanced \underline{D}ynamic \underline{R}elay \underline{P}oint Protocol}
\label{sec:edrp-design}

Aforementioned key insights (Section-\ref{sec:insights}) and design constraints (Section-\ref{sec:theory}) highlight the need for enhancing DRP to ensure its suitability for real-world applications. 
So, we developed EDRP comprising of three components: (i)~DRP \cite{guo2023rateless} (ii)~Link-quality aware CSMA, and (iii)~Machine Learning-based block size selection for rateless codes.

\vspace{-0.10in}
\subsection{LQ-CSMA: \underline{L}ink-\underline{Q}uality-aware \underline{CSMA}}
\label{subsec:lqcsma-design}

DRP's inefficiency in isolating the transmissions of multiple senders 
affects its performance. These transmissions become a mere broadcast of rateless packets whose channel access is governed by CSMA. When CSMA handles this mere broadcast of rateless packets, a highly connected node may receive a higher back-off delay and vice versa. CSMA does not follow the principle of DRP to always set lower back-off delays for highly connected nodes because CSMA randomly chooses a back-off delay value from a larger range.
LQ-CSMA combats this by logically restricting the size of the back-off range. This logically restricted range is dynamically placed towards the lowest (or) highest back-off value of the actual larger range, based on a node's average LQ with its children. 

\begin{figure*}[t]
    \centering
    \includegraphics[width=\linewidth, keepaspectratio]{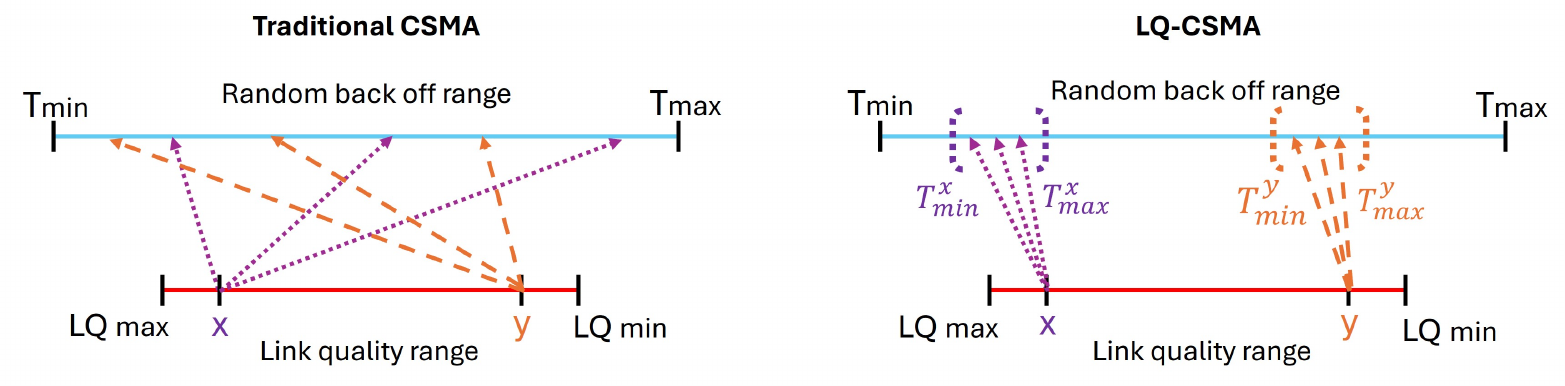}
    \caption{Illustration of the difference between traditional CSMA and LQ-CSMA}
    \label{fig:LQ_CSMA}
\end{figure*}

\textbf{\textit{Motivation.}} From the previous section, it is known that the probability of DRP perfectly isolating the transmissions of multiple senders is very low. This is due to the fact that multiple senders in the same neighborhood experience similar link quality conditions with their children. The log files show that only 
9\%,11\%
of the disseminations are 
perfectly, partially
isolated by the DRP's delay timer, respectively. So, the rest of the transmissions are going to overlap while the channel access is governed by CSMA. During the transmission of bulk data using rateless codes, bulk data of 1000 bytes will be chunked into smaller blocks. There will be multiple larger packets comprising several blocks. These packets are continuously transmitted. Hence, there will be a lot of activity in the channel. Elevated channel activity triggers CSMA random-backoff delays and collisions, as multiple nodes continuously transmit large packets, contrary to the usual data traffic seen in IoT applications. The principle of DRP\cite{guo2023rateless} is to let the highly connected nodes to transmit the bulk data first, by setting a lower delay for them to start transmissions. This principle is not satisfied when traditional CSMA is used. This motivates us to develop a Link-Quality-aware CSMA that assigns random back-off timers based on the average link quality between a sender and its children.

The working of traditional CSMA \cite{7460875}
and LQ-CSMA is illustrated on the left and right sides of Figure~\ref{fig:LQ_CSMA} respectively. In Figure~\ref{fig:LQ_CSMA}, the top, long blue line denotes CSMA's random back-off range, and the bottom, short red line denotes the link-quality estimation range. $T_{min}$ and $T_{max}$ denote the lower and upper limits of CSMA's random back-off range. $LQ_{max}$ and $LQ_{min}$ denote the upper and lower limits of the link quality estimation range. For this illustration, we assume that $LQ_{max}$, $LQ_{min}$ are the best and worst link qualities, respectively. With this assumption, \textit{x} and \textit{y} shown on the link quality range are high and low-quality links, respectively. Traditional CSMA randomly chooses a back-off delay for both \textit{x} and \textit{y}. This random value can be either a lower (or) higher delay. If traditional CSMA chooses a higher delay for \textit{x} and a lower delay for \textit{y}, this will violate the above-explained principle of DRP. According to the principle of DRP, \textit{x} should be assigned a lower back-off delay since it is a high-quality link and \textit{y} should be assigned a higher back-off delay since it is a low-quality link. However, the traditional CSMA violates this principle as it randomly chooses a back-off delay value from the larger range [$T_{min}$, $T_{max}$].

LQ-CSMA combats this by limiting the ranges for different link qualities as shown on the right side of Figure \ref{fig:LQ_CSMA}. For a high-quality link, like \textit{x}, LQ-CSMA restricts the range to be [$T^{x}_{min}$, $T^{x}_{max}$] in such a way that this range is closer to $T_{min}$, leading to a lower delay for high-quality links. Similarly, for a low-quality link, like \textit{y}, LQ-CSMA forces the range to be [$T^{y}_{min}$, $T^{y}_{max}$] such that it falls closer to $T_{max}$, leading to higher delays for low-quality links. Our experiment logs indicated that LQ-CSMA suffered a slight goodput degradation because of two cases where LQ-CSMA provides (i) a few completely overlapping back-off ranges when two nodes have the same average LQ with their children and, (ii) a lot of partially overlapping back-off ranges when two nodes have closer average LQ with their children. While the completely overlapping ranges are solved by choosing a random-value within the window.
The partially overlapping range issue is solved by optimizing the input parameters (X,Y) of Alg. \ref{alg:lq_csma}.

LQ-CSMA is described in Algorithm \ref{alg:lq_csma}. This explains the dynamic back-off range selector algorithm based on the average link quality of the sender with its children. This algorithm takes the inputs: (i) L - current average link quality estimation of the sender with its children, 
(ii) $LQ_{min}$~-~The lower limit of the link-quality estimation range, (iii) $LQ_{max}$~-~Upper limit of the link quality estimation range, (iv) $T_{min}$~-~The lower limit of the CSMA back-off range, (v) $T_{max}$~-~The upper limit of the CSMA back-off range, (vi) X~-~A numeric constant to select the lower limit of the new range, (vi) Y~-~A numeric constant to select the upper limit of the new range. With these input values, the LQ-CSMA algorithm maps the input L to a value in the larger actual back-off range of traditional CSMA. This mapping will make sure that high and low-quality links are mapped to smaller and higher back-off delays respectively. Using this mapped value \textit{val} and the inputs \textit{X,Y}, a smaller logical range is selected from the actual larger range. Finally, a random value is chosen from this smaller logical range, that abides by the principle of DRP.
\begin{algorithm}[t!]
\caption{LQ-CSMA dynamic back-off range selector}
\label{alg:lq_csma}
\begin{algorithmic}
\State Input: L, $LQ_{min}$, $LQ_{max}$, $T_{min}$, $T_{max}$, X, Y 
\State $val \gets T_{min} + \frac{L - LQ_{min}}{LQ_{max} - LQ_{min}} * (T_{max} - T_{min})$
\State $range \gets [val-X, val+Y]$
\State return range
\end{algorithmic}
\end{algorithm}
\vspace{-0.15in}
\subsection{ML-based block size selection algorithm for rateless coding}
\label{subsec:mlbss-design}

\textbf{\textit{Rateless Codes}} divide the bulk data of 1000 bytes into smaller chunks. These chunks are  XOR'ed with other chunks to form encoded blocks. These encoded blocks are packed into the payload of the packet to be transmitted. Multiple larger packets comprising these blocks are transmitted to the receiver. When the receiver collects enough useful encoded blocks, it can decode the bulk data from these smaller encoded blocks~\cite{du2016rateless}. In general, when a packet is transmitted, the receiver checks the CRC for each packet. If the packet fails the CRC check, the receiver will drop the packet as the failed CRC check means the packet is corrupted and not usable. While transmitting a packet that contains multiple encoded rateless blocks, the CRC check for the packet is disabled. Instead, a CRC check is done for each encoded block in the packet. The main reason to do this is if the packet containing rateless blocks is corrupted, only certain corrupted blocks are not usable but the uncorrupted blocks can be used to decode the bulk data. Block size is chosen at the beginning of the transmission and adapted based on the link quality conditions.

\textbf{\textit{Motivation.}}  The goodput is calculated only on the useful data (encoded blocks) and not on the overhead data like packet headers and CRC bytes. Block size decides the overhead. The number of smaller blocks packed in a packet will be higher than the number of bigger blocks packed in a packet. Since each block is given a CRC byte, a packet containing smaller blocks will have more CRC byte overhead than a packet containing bigger blocks. To obtain higher goodput performance, it is intuitive to reduce the overhead by setting bigger block sizes. However, the block size also affects the goodput performance of the protocol.  During high link quality conditions, packet corruption will be lower. So, using larger block sizes leads to lower overhead and higher goodput, whereas using smaller block sizes at high link quality conditions leads to higher overhead and lower goodput. In the same way, during low link quality conditions, packets experience more corruption. 
So, using smaller block sizes will deliver more useful encoded blocks to the receiver
R1-14
\cite{du2016rateless}.
The link quality in real-world scenarios fluctuate erratically. Usually, the receiver sends a feedback packet on the received blocks. The sender adapts the block size based on this feedback~\cite{du2016rateless}. These feedback packets are not useful in multi-cast scenarios like DRP as the sender has to wait for feedback from multiple nodes, further reducing the goodput. So, DRP uses a passive acknowledgment scheme and hence cannot receive feedback from the receiver to adapt the block size. This motivates us to develop a machine learning-based block size selection algorithm. While the traditional block size selection algorithms use past link quality, obtained via feedback packets, to set a block size for future transmissions, we aim to develop an ML-based block size selection algorithm that \emph{predicts} the future link quality to set the block size for future transmissions. 

\textbf{\textit{Machine Learning Problem Formulation:}} 
The goal of this machine-learning algorithm is to predict a block size that will optimize the goodput for future transmissions. We formulated this problem as an ordinal regression problem that takes in the inputs \textit{Rank}, \textit{Packet Delivery Ratio (PDR)}, \textit{Required Number of Packets (RNP)}~\cite{sundaram2024mars} to output an optimal block size for future transmissions. 

The input vector can be represented as:

\begin{equation}
Input_{i} = [Rank, PDR, RNP]    
\end{equation}

\textit{Rank} is obtained from the RPL protocol and indicates the relative position of a node to the root. PDR and RNP are well-known link quality metrics. RNP \cite{sundaram2024mars} is beneficial for identifying the distribution of underlying packet losses. 

The output vector can be represented as:
\begin{equation}
    Output_{i} = \{i|j|k\}
\end{equation}

where $\{i,j,k\}$ represents smaller, medium, and larger block sizes, respectively.

\textbf{Motivation for ordinal regression.} From a machine-learning perspective, our problem can be framed as a classification task to predict one of three possible classes. Standard classifiers, such as decision trees and SVMs, can be applied to this formulation. However, this straightforward approach overlooks crucial information embedded in the block size values $\{i,j,k\}$ where $i < j < k$. Specifically, in our application, misclassifying a block size of i as k carries a significantly higher cost than misclassifying it as j, and vice versa. Empirically, we notice that using larger block sizes during low-link quality scenarios and smaller block sizes during high-link quality scenarios drastically plummet the goodput performance. This highlights the importance of both relative and absolute ordering between classes. In statistics and machine learning literature, such problems fall under the domain of ordinal regression \cite{pedregosa2017consistency}, specifically designed for classification tasks where labels exhibit a natural order.

\textbf{Loss function.} Many methods exist to deal with ordinal regression and for different types of models. In our problem, the absolute values of block sizes carry important meaning, and we leverage these to derive the following cost-sensitive classification approach: 
 the cost of misclassifying the block size $y_i$ against the other block size $y_j$ is equal to their absolute value difference, $|y_i - y_j|$. 
This incorporates the relative and absolute ordering between classes and can be adapted to standard classification methods. Another possible approach for our ordinal regression problem is to treat it as a regular regression problem but to constrain the model to output only the three possible values of $\{i, j, k\}$. This additional constraint usually requires some modification to standard algorithms/methods in regression. Both of these approaches lead to the same learning algorithm for decision trees optimized with the TAO algorithm.

\begin{table}[]
\tiny
\centering
\caption{Prediction Models and Results}
\label{tab:pred}
\resizebox{\columnwidth}{!}{%
\begin{tabular}{|c|c|c|}
\hline
\begin{tabular}[c]{@{}c@{}}ML Models\\  \&\\ optimizations\end{tabular} &
  \begin{tabular}[c]{@{}c@{}}Training \\ Accuracy \\ (in \%)\end{tabular} &
  \begin{tabular}[c]{@{}c@{}}Testing \\ Accuracy\\ (in \%)\end{tabular} \\ \hline
LR            & 90.08 $\pm$ 0.08 & 90.02 $\pm$ 0.19 \\ \hline
SVM           & 93.77$\pm$0.06   & 93.07$\pm$0.14   \\ \hline
CART    & \textbf{96.94$\pm$1.92}   & 86.45$\pm$0.20   \\ \hline
TAO\cite{carreira2018alternating}-CART    & 93.63$\pm$0.31   & 92.12$\pm$0.26   \\ \hline
TAO\cite{carreira2018alternating}-oblique   & 95.52$\pm$0.08   & \textbf{95.00$\pm$0.34}   \\ \hline
\end{tabular}%
}
\end{table}

\textbf{\textit{Data collection}} experiment was conducted on the deployment shown in Fig. \ref{fig:topo}, with the experimental setting explained in 
Section
\ref{sec:DRP_perf}. We ran multiple experiments by fixing a different block size for each experiment. Finally, we hand-curated a dataset with 90,000 points showing the optimal block size for different link quality conditions.

\textbf{\textit{Models and accuracies:}} We developed SVM, Logistic Regression (LR), CART, and TAO\cite{carreira2018alternating}-optimized decision trees. All these models were trained with cost-sensitive loss for our ordinal regression problem. The training and testing accuracies of different models obtained through 5-fold cross-validation are tabulated in Table \ref{tab:pred}. LR and SVM classifiers perform the worst. CART achieves better accuracy than LR and SVM but the high difference between its training and testing accuracy shows that CART is not well-generalized for unseen data, yielding bad results in real-world deployments \cite{sundaram2025comnets}. TAO~\cite{carreira2018alternating}-CART is the CART tree optimized by the TAO algorithm. TAO-CART optimizes CART well and makes it suitable for real-world deployments as the difference between its training and testing accuracy is lower. TAO optimization for oblique decision tree achieves the most accurate results. TAO-Oblique is also well-optimized for real-world deployments as the difference between its training and testing accuracies is very low. The reasons for this are twofold. First, oblique decision trees use linear (hyperplane) splits at decision nodes, which are more powerful than traditional axis-aligned trees. Second, the more recent TAO algorithm is very effective in optimizing our desired objective function (cost-sensitive loss) and helps to produce more accurate models. TAO-oblique decision trees are much smaller and faster during inference.

\textbf{Novelty of ML-BSS in EDRP:} Unlike prior rateless block size adaptation methods \cite{du2016rateless} that rely on explicit feedback from receivers, DRP operates under a passive acknowledgment scheme, making such feedback unavailable. This renders existing adaptation algorithms inapplicable. To address this challenge, we are, to the best of our knowledge, the first to demonstrate that block size adaptation can be achieved without feedback by predicting future link quality using only locally available link quality estimations. We reformulate the prediction task as an ordinal classification problem, capturing the fact that different block size mispredictions have asymmetric effects on goodput. To align the learning process with system-level performance, we design a custom loss function in which penalties grow with the magnitude of block size deviation. Furthermore, we employ the TAO optimization framework to produce compact, highly accurate decision trees suitable for deployment on TelosB-class motes with strict memory and processing constraints. This combination of feedback-free operation, ordinal regression reformulation, custom loss, and hardware-aware optimization, trained on 90,000 real-world field measurements rather than simulations, makes ML-BSS a domain-specific, system-aligned adaptation mechanism rather than a simple application of standard machine learning models.

\vspace{-0.10in}
\section{Real-world evaluations of EDRP}
\begin{figure}[t]
    \centering
    \includegraphics[width=0.9\linewidth, height=7cm]{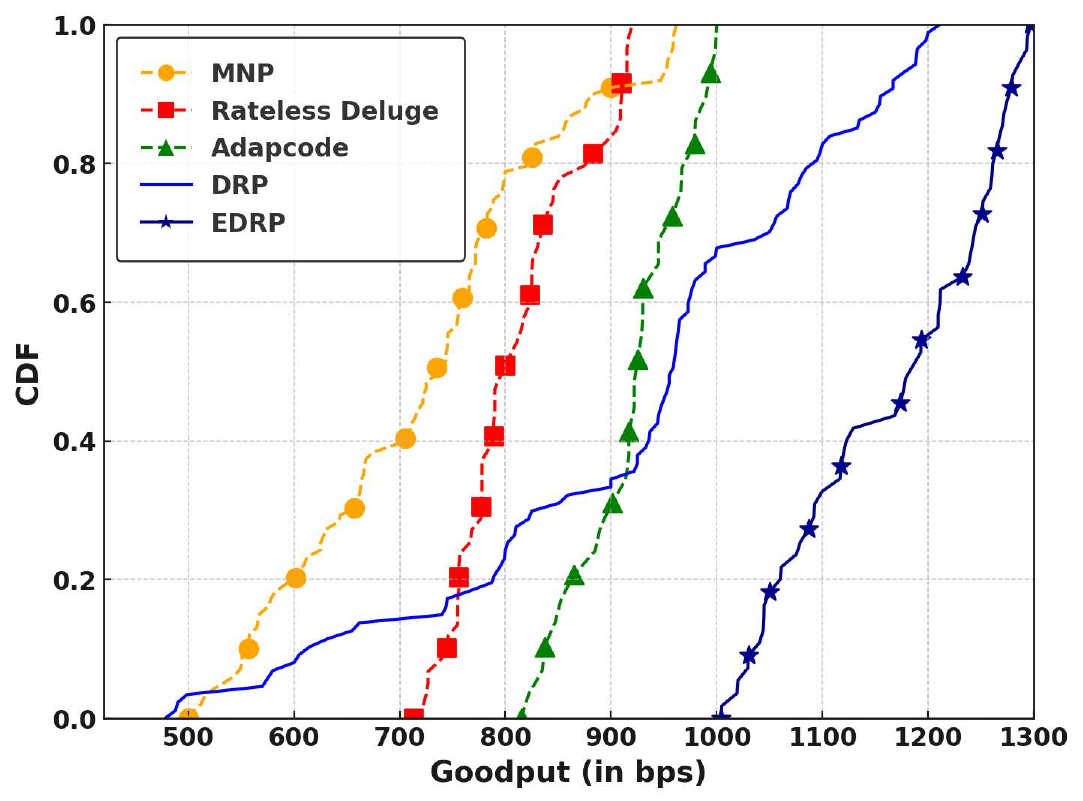}
    \caption{Large-scale evaluations of EDRP}
    \label{fig:eval_cdf}
\end{figure}
In this section, we explain our in-field evaluations of EDRP. EDRP consists of three components: (i) DRP, (ii) ML-based block size selection algorithm (ML-BSS), and (iii) LQ-CSMA. We also conducted component analysis experiments to understand the contributions of each above-mentioned components to the overall goodput performance of EDRP. In total, we transmitted $1.7592 \times 10^{6}$ bytes of data for these experiments.

\begin{figure}[t]
    \centering
    \includegraphics[width=0.9\linewidth, height=7cm]{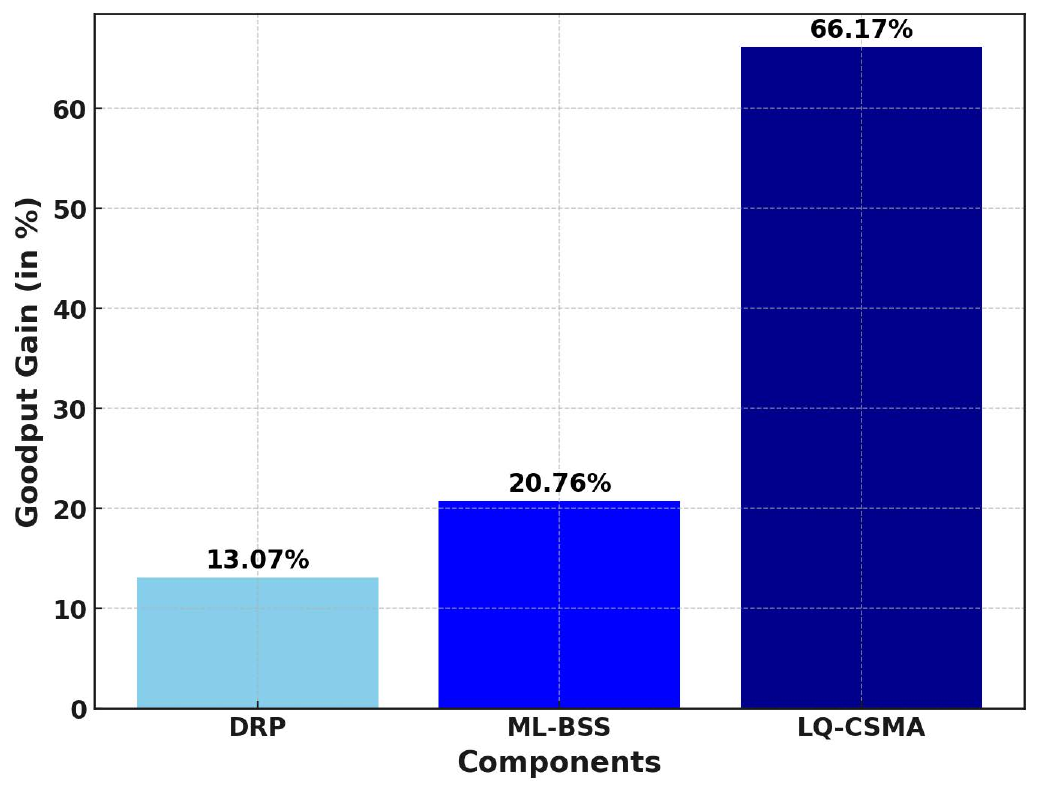}
    \caption{Component analysis of EDRP}
    \label{fig:comp_analysis}
    \vspace{-0.18in}
\end{figure}

Hardware, topology setup, experimental settings, metrics, and benchmarks are the same for this evaluation as explained in 
Section
\ref{sec:DRP_perf}. Our experiments obtained optimal results while setting the input parameters of Algorithm \ref{alg:lq_csma} as $LQ_{min}$=0, $LQ_{max}$=1, $T_{min}$=20, $T_{max}$=640, X=61, and Y=-30.  We arrived at these values using trial and error experimentally, and we obtained very good performance. An optimal value for these input variables will vary depending on the deployed environment and the radio technology used.

Space and Time complexity of ML-BSS:
An important design consideration of any data-driven, ML-based system is the space and time complexity of the ML-model deployed in the system. Since TAO-Oblique is a decision tree, we deployed this as IF…ELSE statements in TelosB motes. This took only 300B of space with a minimal inference latency of 0.007ms.



Figure~\ref{fig:eval_cdf} depicts the overall goodput performance of EDRP. This shows the CDF of EDRP and its competitors, MNP\cite{arakadakis2021firmware}, Rateless Deluge \cite{arakadakis2021firmware}, AdapCode \cite{sundaram2025machine} and DRP \cite{guo2023rateless}. Deep insights into the performance of the competitors have been explained in 
Section
\ref{sec:insights}. It is evident from this graph that EDRP achieves better goodput than the competitors. Especially, EDRP overcomes the inefficiencies of DRP and provides stable performance in real-world deployments, making EDRP easily adoptable for real-world applications. On average, EDRP achieves an overall goodput gain of 39.43\% than its competitors. Specifically, EDRP achieves an average goodput gain of 61.36\%, 43.61\%, 26.67\%, and 26.10\% than MNP\cite{arakadakis2021firmware}, Rateless Deluge \cite{arakadakis2021firmware}, AdapCode \cite{sundaram2025machine} and DRP \cite{guo2023rateless}. 


Table \ref{tab:completion_times_reduction} presents the completion time for disseminating a 1000 Byte payload to the entire network, along with the reduction factors showing how much faster EDRP completes dissemination compared to the other protocols. EDRP achieves the shortest completion time of 6.67s, while the slowest protocol, MNP, requires 12.31s, making EDRP approximately 1.85× faster. Rateless Deluge, AdapCode, and DRP take 1.56×, 1.36×, and 1.29× longer than EDRP, respectively. On average, EDRP completes dissemination 1.39× faster than the other protocols combined. These results underscore that the goodput improvements reported earlier directly translate into substantial reductions in end to end dissemination time, demonstrating the practical advantage of EDRP in real network deployments.

\begin{figure}[t]
    \centering
    \includegraphics[width=0.9\linewidth, height=7cm]{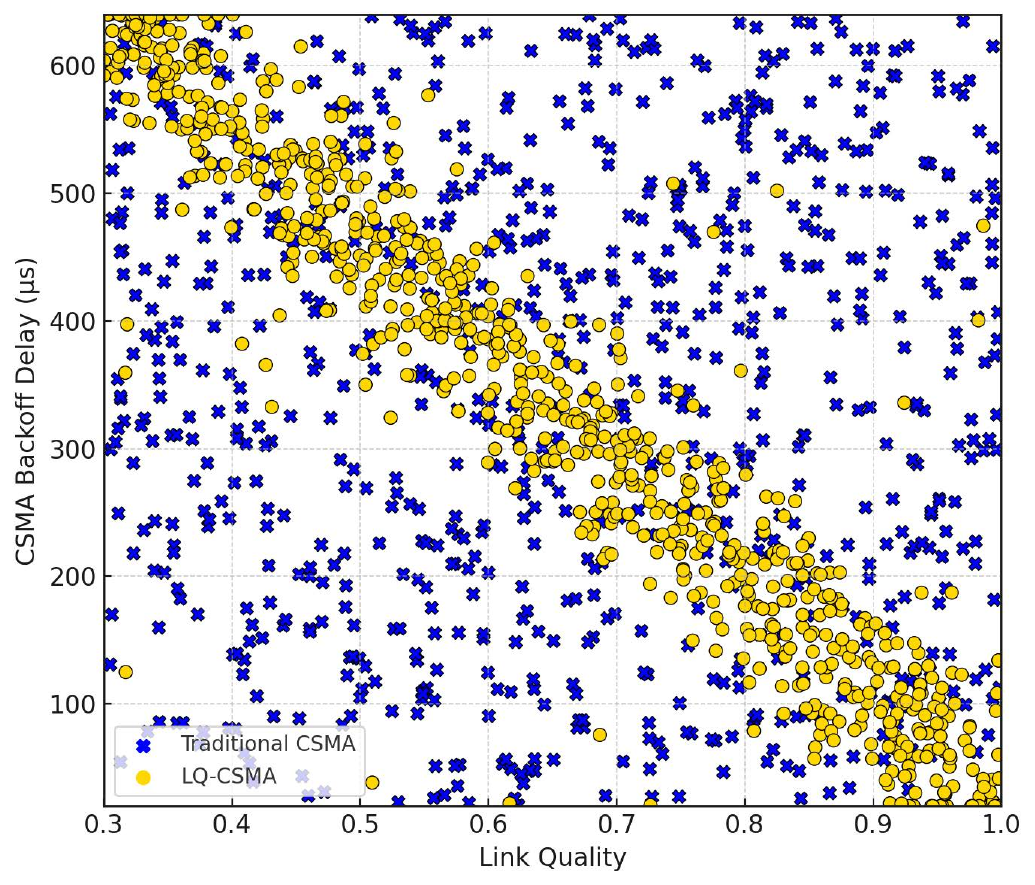}
    \caption{Backoff delay comparison: CSMA vs. LQ-CSMA}
    \label{fig:eval-csma}
\end{figure}

\begin{table}[t]
\centering
\caption{Estimated completion times for 1000~B dissemination and reduction factors of EDRP against other protocols}
\resizebox{\columnwidth}{!}{
\begin{tabular}{lcc}
\hline
\textbf{Protocol} & \textbf{Completion Time (in s)} & \textbf{EDRP's reduction factor} \\
\hline
MNP\cite{arakadakis2021firmware} & 12.31 & $12.31 / 6.67 \approx 1.85\times$ \\
Rateless Deluge\cite{arakadakis2021firmware} & 10.39 & $10.39 / 6.67 \approx 1.56\times$ \\
AdapCode\cite{sundaram2025machine} & 9.09 & $9.09 / 6.67 \approx 1.36\times$ \\
DRP\cite{guo2023rateless} & 8.60 & $8.60 / 6.67 \approx 1.29\times$ \\
Average of others & 9.29 & $9.29 / 6.67 \approx 1.39\times$ \\
EDRP & 6.67 & --- \\
\hline
\end{tabular}}
\label{tab:completion_times_reduction}
\vspace{-0.3in}
\end{table}

Figure \ref{fig:comp_analysis} shows the component analysis of EDRP. We conducted an extensive experimental campaign for this analysis. The components DRP, ML-BSS, and LQ-CSMA were run stand-alone. While running ML-BSS, it was just a broadcast of rateless packets using the ML-BSS algorithm. While running LQ-CSMA, it was just the broadcast of rateless packets of fixed block size without the ML-BSS algorithm. This figure shows that DRP\cite{guo2023rateless}, ML-BSS, and LQ-CSMA contribute 13.07\%, 20.76\%, and 66.17\% to EDRP's goodput gain respectively. This figure shows that LQ-CSMA and ML-BSS alone can yield 86.93\% of the overall goodput gain than the competing protocols. 

\textbf{Delay timer comparison of traditional CSMA and LQ-CSMA:} LQ-CSMA takes the link quality into consideration to prioritize high link quality nodes with lower back-off delay and penalize low link quality nodes with higher back-off delay. We evaluated this by plotting the back-off delay experienced by the packets under traditional CSMA and LQ-CSMA algorithms and proved that LQ-CSMA works as intended.

Figure~\ref{fig:eval-csma} shows the relationship between link quality and 
CSMA backoff delay for Traditional CSMA and LQ-CSMA. In Traditional CSMA, back-off delays are uniformly distributed across the link quality range, with no observable trend (Pearson correlation \cite{benesty2018pearson} = -0.032). This confirms that the traditional CSMA mechanism assigns delays randomly, irrespective of link quality. In contrast, LQ-CSMA exhibits a strong negative correlation (Pearson correlation \cite{benesty2018pearson} = -0.947), generally assigning shorter delays to higher-quality links and longer delays to lower-quality links. This prioritization bias helps reduce collisions in contention domains, though occasional deviations occur due to the non-deterministic nature of CSMA, where channel occupancy can allow lower-quality links to transmit earlier in certain cases.

\section{Conclusion}
We conducted an extensive, in-field performance analysis of DRP and found that it is inefficient in handling real-world fluctuations and does not fit well for real-world deployments. Our in-depth analysis of DRP's transmission events showed that the reason for DRP's poor performance stems from its delay timer, as it isolates only 9\% of the transmissions and the rest becomes a mere broadcast of rateless packets. To ameliorate this issue, we developed EDRP that comprises: (i) DRP, (ii) ML-BSS, and (iii) LQ-CSMA. LQ-CSMA combats the real-world link quality fluctuations and assigns constrained random back-off delay to preserve the principle of DRP. ML-BSS predicts the future link quality and adopts the rateless block size for future transmissions. EDRP achieves an average throughput gain of 39.43\% than the competing protocols.

\section{Discussion and Future Work}
We acknowledge that the proposed LQ‑CSMA algorithm, while designed to bias channel access in favor of higher‑quality links, does not guarantee strict transmission ordering in all scenarios. The algorithm reduces the back‑off delay proportionally to link quality, which in expectation increases the likelihood that high‑quality links transmit earlier. However, the inherent randomness in traditional CSMA and the stochastic nature of wireless channel availability can lead to situations where this priority is not fully preserved.

For example, consider two nodes, A and B, both ready to transmit. If A has a higher link quality, it may be assigned a shorter back-off time x, while B waits for a longer duration y, such that y=3x. However, if the channel is busy during A's first three transmission attempts and only becomes free afterward, both A and B may attempt to transmit simultaneously, resulting in a collision. Furthermore, depending on the ratio of x and y, it is also possible for B to complete its back-off earlier than A and transmit first. This example suggests that the proposed approach does not always prioritize high-quality links as intended. These effects are a fundamental property of contention‑based access mechanisms and cannot be completely eliminated without introducing a more rigid scheduling structure, such as TDMA. In practice, our in‑field experiments show that LQ‑CSMA still significantly reduces the probability of collisions among nodes with similar link quality, thereby improving goodput compared to DRP’s original CSMA behavior. Nonetheless, we recognize that residual collisions are inevitable under contention‑based MACs. Future work could explore hybrid approaches that combine LQ‑CSMA with lightweight coordination or reservation schemes to further reduce inversion events and contention‑induced collisions.

\balance

\bibliographystyle{IEEEtran}
\bibliography{ref}
\end{document}